\documentclass[a4paper,times,amssymb,nofootinbib]{article}
\usepackage[margin=2cm]{geometry}
\usepackage{authblk}
\usepackage{xcolor}
\usepackage{hyperref}
\hypersetup{colorlinks=true, linkcolor=blue, urlcolor=blue, citecolor=blue}
\usepackage{soul}
\usepackage[figuresright]{rotating}
\usepackage{mathtools}
\usepackage{amsmath}
\usepackage{graphicx}
\usepackage{dcolumn}
\usepackage{bm}
\usepackage{url}
\usepackage[bottom]{footmisc}

\begin{document}

\title{tParton: Implementation of next-to-leading order evolution of transversity parton distribution functions}

%% use optional labels to link authors explicitly to addresses:
\author[1,*]{Congzhou M Sha}
\affil[1]{\textit{Penn State College of Medicine, 500 University Dr, Hershey, PA, 17033}}
\author[2]{Bailing Ma}
\affil[2]{\textit{Wake Forest University School of Medicine, 475 Vine St, Winston-Salem, NC 27101}}
\affil[*]{\textit{Corresponding author. Email: \href{mailto:cms6712@psu.edu}{cms6712@psu.edu}}}
\date{}
%%\author{Congzhou M Sha}
%%\email{cms6712@psu.edu}
%%\author{}
%%
%%\address{}

\maketitle
\begin{abstract}
We provide code to solve the Dokshitzer–Gribov–Lipatov–Altarelli–Parisi (DGLAP) evolution equations for the transversity parton distribution functions (PDFs), which encode the hadron's transverse spin structure. PDFs describe the probability of finding a parton within a hadron moving with a certain light-cone momentum fraction $x$ with respect to the hadron. The transversity PDF is the difference in a transversely polarized hadron between the number density of partons with momentum fraction $x$ and spin parallel to that of the hadron and the number density of partons with the same momentum fraction and spin antiparallel to the hadron. Though codes are widely available for the evolution of unpolarized and polarized PDFs, there are few codes publicly available for the transversity PDF. Here, we present Python code which implements two methods of solving the leading order (LO) and next-to-leading order (NLO) approximations of the DGLAP equations for the transversity PDF, and we highlight the theoretical differences between the two.
\end{abstract}


%% MSC codes here, in the form: \MSC code \sep code
%% or \MSC[2008] code \sep code (2000 is the default)

%%
%% Start line numbering here if you want
%%
% \linenumbers

%% main text

\section{Introduction}
The distribution of quarks and gluons inside hadrons can be described by parton distribution functions (PDFs)~\cite{altarelli}. In the parton picture, a PDF describes the probability of finding a quark or a gluon in a fast-moving hadron with a certain fraction of the light-front momentum of the parent hadron. The knowledge of PDFs is crucial for our understanding of quantum chromodynamics (QCD) and for the interpretation of high-energy experiments involving hadrons, and there has been increasing interest, both in theory and in experiment, in the nucleon's transversity PDF~\cite{Ledwig:2010tu, Bacchetta:2012ty, Sharma:2023wha, Wakamatsu:2008ki, Cloet:2007em, Anselmino:2008jk, Anselmino:2013vqa, Kang:2015msa, Radici:2015mwa}. 
\par For partons moving in a collinear direction with the parent hadron, the nucleon's spin structure at leading twist can be described by three independent PDFs: the unpolarized distribution, $q(x)$, the helicity distribution, $\Delta q(x)$, and the transversity distribution, $\Delta_T q(x)$. Experimentally, the transversity distribution is the least known, since it can only be measured in processes involving two hadrons due to the processes' chiral-odd property, such as in semi-inclusive deep inelastic scattering. The analysis of experimental data is also difficult since it involves transverse momentum dependent (TMD) PDFs and their QCD evolution~\cite{Sharma:2023wha}. In fact, the transversity distribution was extracted from experiments for the first time in 2008~\cite{Anselmino:2008jk}.
\par The calculation of PDFs is usually performed at a fixed energy. In order to compare predictions among different experiments and theoretical calculations, it is necessary to evolve the PDF to a shared energy scale. The unpolarized and helicity distributions have been extensively studied for many years, both experimentally and theoretically, and codes to perform their evolutions are widely available, such as the QCDNUM~\cite{qcdnum}, EKO~\cite{EKO}, mcEvol~\cite{Hautmann:2017xtx, Hautmann:2017fcj} and HOPPET~\cite{Salam:2008qg} packages.
\par In contrast, while the theoretical framework for transversity PDF evolution was established in the late 1990s~\cite{Hayashigaki97, Vogelsang97, hirai}, the implementation provided by Hirai et al. was written in Fortran and is now nearly 30 years old~\cite{hirai}. Furthermore, the code is no longer accessible, as the link referenced in Hirai et al.~\cite{hirai} is inactive. APFEL++~\cite{Bertone:2013vaa,Bertone:2017gds}, a C++ package, offers another implementation of transversity evolution, employing a method similar to Hirai's by numerically solving the DGLAP equation in $t$-steps. Alternatively, one can solve the equation in Mellin space and reconstruct the evolved PDF by inverse Mellin transform. Here, we present a transversity PDF evolution code package that, for the first time, incorporates both methods simultaneously. The method presented by Hirai et al.~\cite{hirai} can be computationally expensive and more discretization-dependent compared to Mellin moment method proposed by Vogelsang~\cite{Vogelsang97}.
\par In this work, we (1) use Mathematica~\cite{mathematica} to verify the correctness of the splitting function Mellin moments given by Vogelsang, (2) provide both Mathematica and Python implementations of the method used by Hirai et al.~\cite{hirai}, and (3) provide a Python implementation of the method proposed by Vogelsang~\cite{Vogelsang97}. Although the transversity anomalous dimension was recently calculated up to 3-loop order~\cite{Blumlein:2021enk}, here we only implement the evolution up to NLO. A discrepancy exists between the two methods. We discuss the advantages and disadvantages of choosing one method over the other. Additionally, we show a comparison of the results of our code versus that of APFEL++, as well as the dependence of the Hirai method on discretization effects. The Python implementation is called tParton, and is available on the Python Package Index {\color{blue}\href{https://pypi.org/project/tparton/}{https://pypi.org/project/tparton/}} with source code on GitHub ({\color{blue} \href{https://github.com/mikesha2/tParton/}{https://github.com/mikesha2/tParton/}}). The package can be installed using the command: \verb|pip install tparton|.

\section{Methods}
\subsection{Defining equations for NLO evolution of the transversity PDF}
We start with the DGLAP equation for evolution of the transversity PDF, which is Eq. (2.6) of \cite{hirai}:
\begin{equation}
	\label{eq:dglap}
	\frac{\partial}{\partial t}\Delta_T\tilde{q}^\pm(x,t)=\frac{\alpha_s(t)}{2\pi}\Delta_T\tilde{P}_{q^\pm}(x)\otimes\Delta_T\tilde{q}^\pm(x,t)
\end{equation}
where $t:=\ln Q^2$, $Q^2$ is the energy scale of the PDF (e.g. the dimuon-mass squared in the Drell-Yan process),
\begin{equation}
	\tilde f(x)=xf(x),
\end{equation}
\begin{equation}
	\label{eq:conv}
	f(x)\otimes g(x):=\int_x^1\frac{dy}y f\left(\frac xy\right)g(y),
\end{equation}
the NLO $\alpha_S$ is given by~\cite{hirai,Vogelsang97}
\begin{equation}
	\label{eq:alpha}
	\alpha_S^{NLO}(Q^2)=\frac{4\pi}{\beta_0\ln{\left(\frac{Q^2}{\Lambda^2}\right)}}\left[1-\frac{\beta_1\ln{\left(\ln{\left(\frac{Q^2}{\Lambda^2}\right)}\right)}} {\beta_0^2\ln{\left(\frac{Q^2}{\Lambda^2}\right)}}\right],
\end{equation}
where $\Lambda$ is the QCD scale parameter, $\beta_0=\frac{11}3C_G-\frac43T_RN_f$, $\beta_1=\frac{34}3C_G^2-\frac{10}3C_GN_f-2C_FN_f$, $C_G=N_c$, $C_F=\frac{N_c^2-1}{2N_c}$, $T_R=\frac12$, $N_c=3$ is the number of colors, and $N_f$ is the number of flavors. The LO approximation of $\alpha_S$ is obtained by setting $\beta_1=0$. $\Delta_T P_{q^\pm}$ is known as the transversity splitting function.

\par Note that the convolution in Eq. (\ref{eq:conv}) is symmetric under interchange of $f$ and $g$, with the substitution $z=\frac xy$, $dz=-\frac{x}{y^2}dy$:
\begin{eqnarray}
	\int_x^1\frac{dy}{y}f\left(\frac xy\right)g(y)=\int_1^x-\frac{y^2 dz}{x}f(z)g\left(\frac xz\right)=\int_x^1\frac{dz}{z}g\left(\frac xz\right)f(z).
\end{eqnarray}
Also note that although Eq.~(\ref{eq:conv}) was originally defined for the non-tilde equation,
\begin{equation}
	\frac{\partial}{\partial t}\Delta_T q^\pm(x,t)=\frac{\alpha_s(t)}{2\pi}\Delta_T P_{q^\pm}(x)\otimes\Delta_T q^\pm(x,t),
\end{equation}
the tilde function satisfies the same form due to the following:
\begin{align}
	\label{eq:tildeform}
	\frac{\partial}{\partial t}\Delta_T\tilde{q}^\pm(x,t)=\frac{\partial}{\partial t}x \Delta_T q^\pm(x,t)
	&=\frac{\alpha_s(t)}{2\pi}x\int_x^1\frac{dy}{y}\Delta_T P_{q^{\pm}}\left(\frac{x}{y}\right)\Delta_T q^{\pm}(y,t)\notag\\
	&=\frac{\alpha_s(t)}{2\pi}\int_x^1\frac{dy}{y}\frac{x}{y}\Delta_T P_{q^{\pm}}\left(\frac{x}{y}\right)y\Delta_T q^{\pm}(y,t)\notag\\
	&=\frac{\alpha_s(t)}{2\pi}\Delta_T\tilde{P}_{q^\pm}(x)\otimes\Delta_T\tilde{q}^\pm(x,t).
\end{align}
\par The full splitting function at NLO is:
\begin{equation}
	\Delta_TP_{q^\pm}(x)=\Delta_TP_{qq}^{(0)}(x)+\frac{\alpha_s(Q^2)}{2\pi}\Delta_TP^{(1)}_{q^\pm}(x),
\end{equation}
with the LO splitting function given by
\begin{equation}
	\label{eq:LO}
	\Delta_TP_{qq}^{(0)}(x)=C_F\left[\frac{2x}{(1-x)_+}+\frac32\delta(1-x)\right],
\end{equation}
where $\delta$ is the Dirac delta function and the plus distribution is defined in the usual way as
\begin{equation}
	\label{eq:plus}
	\int_0^1 dx\ f(x)\left(g(x)\right)_+:=\int_0^1 dx\ \bigg[f(x)-f(1)\bigg]g(x).
\end{equation}
The NLO contribution is given by
\begin{equation}
	\Delta_TP^{(1)}_{q^\pm}(x)=\Delta_TP_{qq}^{(1)}(x)\pm\Delta_TP_{q\bar q}^{(1)}(x),
\end{equation}
with\footnote{A factor of $t$ is erroneously included in the penultimate line of Eq. (A.10) in Ref.~\cite{hirai}, which is not present in the corresponding ArXiv preprint.}
%\begin{widetext}
	\begin{multline}
		\label{eq:NLO}
		\Delta_TP_{qq}^{(1)}(x)=
		C_F^2\bigg[1-x-\left(\frac32+2\ln(1-x)\right)\ln x\frac{2x}{(1-x)_+}
		+\left(\frac38-\frac12\pi^2+6\zeta(3)\right)\delta(1-x)\bigg]\\
		+\frac12C_FC_G\bigg[-(1-x)+\bigg(\frac{67}9+\frac{11}3\ln x+\ln^2 x-\frac13\pi^2\bigg)\frac{2x}{(1-x)_+}
		+\bigg(\frac{17}{12}+\frac{11}{9}\pi^2-6\zeta(3)\bigg)\delta(1-x)\bigg]\\
		+\frac23C_FT_RN_f\bigg[\bigg(-\ln x-\frac53\bigg)\frac{2x}{(1-x)_+}
		-\bigg(\frac14+\frac13\pi^2\bigg)\delta(1-x)\bigg],
	\end{multline}
%\end{widetext}
\begin{equation}
	\Delta_TP_{q\bar q}^{(1)}(x)=C_F\left(C_F-\frac{C_G}{2}\right)\left[-(1-x)+2S_2(x)\frac{-2x}{(1+x)}\right],
\end{equation}
\begin{equation}
	S_2(x)=\int_{\frac{x}{1+x}}^{\frac{1}{1+x}}\frac{dz}z\ln\frac{1-z}z=S\left(\frac{x}{1+x}\right)-S\left(\frac{1}{1+x}\right)-\frac12\left[\ln^2\frac{1}{1+x}-\ln^2\frac{x}{1+x}\right],
\end{equation}
\begin{equation}
	S(x)=\int_x^1dz\ \frac{\ln z}{1-z},
\end{equation}
$\zeta(s)=\sum_{n=1}^\infty\frac{1}{n^s}$ is the Riemann zeta function, and $S(x)$ is known as Spence's function. In Mathematica, the definition of Spence's function is  (via the \textbf{PolyLog} function)
\begin{equation}
	S(z)=-\text{\textbf{PolyLog}}[2,1-z]=-\text{Li}_2(1-z)
\end{equation} 
The dilogarithm itself $\text{Li}_2(z)=-\int_0^z \frac{du}{u}\ln(1-u)$ is also called Spence's function in the literature.

\subsection{Plus distribution and convolution}
According to Eq. (16) of \cite{Ellis:1996nn}, the definition of the Mellin convolution is
\begin{equation}
	\label{eq:newconv}
	(f\otimes g)(x)=\int_0^1\int_0^1\ dy\ dz\ f(y)\ g(z)\ \delta(x-yz)
\end{equation}
Note that this definition of the convolution is manifestly symmetric under interchange of $y$ and $z$. In simple cases, Eq. (\ref{eq:newconv}) reduces to Eq. (\ref{eq:conv}). However, there is a plus distribution regularization prescription in Eqs. (\ref{eq:LO}) and (\ref{eq:NLO}) which must be taken into account. Combining our definitions of the plus distribution in Eq. (\ref{eq:plus}) and the new definition of convolution in Eq. (\ref{eq:conv}), we have
\begin{eqnarray}
	\nonumber
	&&\left(f\otimes (g\cdot h_+)\right)(x)
	\nonumber
	\\&&=\int_0^1 dy\int_0^1 dz\ f(y)g(z)\left(h(z)\right)_+\delta(x-yz)
	\nonumber
	\\&&=\int_0^1 dy\int_0^1 dz\ f(y)\bigg[g(z)-g(1)\bigg]h(z)\delta(x-yz)
	\nonumber
	\\&&=\int_x^1\frac{dz}{z}f\left(\frac xz\right)\bigg[g(z)-g(1)\bigg]h(z),
\end{eqnarray}
matching the prescription given in the QCDNUM documentation~\cite{qcdnum}.
\subsection{Tilde}
\vspace{-1cm}
\begin{eqnarray}
	\label{eq:code}
	\partial_t \tilde f(x)&&=x\cdot\partial_t f(x)
	\nonumber
	\\&& =x\cdot (f\otimes g)(x)
	\nonumber
	\\&&=x\int dy\ dz\ f(y)g(z)\delta(x-yz)
	\nonumber
	\\&&=\int dz\frac{x}{z}f\left(\frac xz\right)g(z)
	\nonumber
	\\&&=\int dz \tilde f\left(\frac xz\right)g(z)
\end{eqnarray}
Note here that we do not have an overall $\frac{1}{z}$ factor in the integrand and the $g$ function is non-tilded. In our code, we implement the final line of Eq. (\ref{eq:code}) instead of Eq. (\ref{eq:tildeform}), with $\tilde f(x)=x \Delta_Tq(x)$ and $g$ as our splitting function.

\subsection{Solution via the convolution theorem for Mellin transforms}
The Mellin transform of a function $f$ is defined as
\begin{equation}
	\label{eq:mellin}
	\mathcal{M}[f](s)=\int_0^\infty x^{s-1}f(x)\ dx.
\end{equation}
For a function $f$ with Mellin transform $\mathcal{M}[f]=\hat f$,
\begin{equation}
	\label{eq:mellininversion}
	f(x)=\mathcal{M}^{-1}[\hat f](x)=\frac{1}{2\pi i}\int_{c-i\infty}^{c+i\infty}x^{-s}\hat f(s) ds,
\end{equation}
where $c$ is any real number for which the integral converges absolutely~\cite{cohen}. When we have a plus distribution which includes factors of $\ln(1-x)$, the Mellin transform of the plus distribution is regularized by Eq. (29) in Vermaseren~\cite{vermaseren}\footnote{There is a typo in the ArXiv version of \cite{vermaseren}, with a missing factor of $x^m$ in the last line of Eq. (29).}:
\begin{equation}
	\mathcal{M}\left[\ln(1-x)^kf_+(x)g(x)\right](s)=\int_0^1 dx\ x^{s-1}\ln(1-x)^k\left(f(x)-f(1)\right)g(x).
\end{equation}
Otherwise, the normal definition of the plus distribution in Eq. (\ref{eq:plus}) applies to Eq. (\ref{eq:mellin}). Note also that~\cite{vermaseren} uses a definition of the Mellin transform which is shifted by $1$ as compared to this work: $m=s-1$. 
The well-known convolution theorem also applies to the Mellin transform~\cite{handbook}:
\begin{equation}
	\mathcal{M}\left[f\otimes g\right]=\mathcal{M}[f]\mathcal{M}[g].
\end{equation}
Consequently, the solution to the DGLAP equation at NLO is such that the moments of the resulting PDF are given by:
%\begin{widetext}
	\begin{multline}
		\label{eq:moment}
		\mathcal{M}[\Delta_Tq^{\pm}](Q^2;s)=\left(1+\frac{\alpha_S(Q_0^2)-\alpha_S(Q^2)}{\pi\beta_0}\left[\mathcal{M}[\Delta_T P_{qq,\pm}^{(1)}](s)-\frac{\beta_1}{2\beta_0}\mathcal{M}[\Delta_T P_{qq}^{(0)}](s)\right]\right)\\\times \left(\frac{\alpha_S(Q^2)}{\alpha_S(Q_0^2)}\right)^{-2\mathcal{M}[\Delta_T P_{qq}^{(0)}](s)/\beta_0}\mathcal{M}[\Delta_T q^{\pm}](Q_0^2;s),
	\end{multline}
%\end{widetext}
which appears as Eq. (20) in Vogelsang \cite{Vogelsang97}. The LO solution is given by~\cite{altarelli}:
\begin{equation}
	\mathcal{M}[\Delta_Tq^{\pm}](Q^2;s)=\left(\frac{\alpha_S(Q^2)}{\alpha_S(Q_0^2)}\right)^{-2\mathcal{M}[\Delta_TP_{qq}^{(0)}](s)/\beta_0}\mathcal{M}[\Delta_T q^{\pm}](Q_0^2;s).
\end{equation}

\par In this solution, the Mellin moments of the evolved distribution $\mathcal{M}[\Delta_T q^{\pm}](Q^2;s)$ are given in terms of the Mellin moments of the initial distribution $\mathcal{M}[\Delta_T q^{\pm}](Q_0^2,s)$, the Mellin moments of the LO and NLO splitting functions ($\mathcal{M}[\Delta_T P_{qq}^{(0)}]$ and $\mathcal{M}[\Delta_T P_{qq,\pm}^{(1)}]$ respectively), and the strong coupling constants at the initial and evolved scales $\alpha_S(Q_0^2)$ and $\alpha_S(Q^2)$.
\par The analytic continuations of the splitting function Mellin moments are given by \cite{Vogelsang97,gluck,moment}:
\begin{equation}
	\label{eq:LO_mom}
	\mathcal{M}[\Delta_T P_{qq}^{(0)}](s)=C_F\left(\frac32-2S_1(s)\right),
\end{equation}
%\begin{widetext}
	\begin{multline}
		\label{eq:NLO_mom}
		\mathcal{M}[\Delta_T P_{qq,\eta}^{(1)}](s)=C_F^2\bigg[\frac38+\frac{1-\eta}{s(s+1)}-3S_2(s)-4S_1(s)\left(S_2(s)-S'_2\left(\eta,\frac{s}{2}\right)\right)
		-8\tilde S(\eta,s)+S'_3\left(\eta,\frac{s}{2}\right)\bigg] \\
		+\frac12 C_FN_C\bigg[\frac{17}{12}-\frac{1-\eta}{s(s+1)}-\frac{134}{9}S_1(s)+\frac{22}{3}S_2(s)
		+4S_1(s)\left(2S_2(s)-S'_2\left(\eta,\frac{s}{2}\right)\right)+8\tilde S(\eta,s)-S'_3\left(\eta,\frac{s}{2}\right)\bigg]\\
		+\frac{2}{3}C_FT_f\left[-\frac14+\frac{10}{3}S_1(s)-2S_2(s)\right],
	\end{multline}
	where
%\end{widetext}
\begin{equation}
	S_1(s)=\gamma+\psi^{(0)}(s+1),
\end{equation}
\begin{equation}
	S_2(s)=\zeta(2)-\psi^{(1)}(s+1),
\end{equation}
\begin{equation}
	S_3(s)=\zeta(3)+\frac12\psi^{(2)}(s+1),
\end{equation}
\begin{equation}
	S'_{\eta,k}(s)=\frac{1}{2}\left(1+\eta^s\right)S_k\left(\frac{s}{2}\right)+\frac12\left(1-\eta^s\right)S_k\left(\frac{s-1}{2}\right),
\end{equation}
\begin{equation}
	\tilde S(\eta,s)=-\frac58\zeta(3)+\eta^s\Bigg[\frac{S_1(s)}{s}-\frac{\zeta(2)}{2}\left(\psi^{(0)}\left(\frac{s+1}{2}\right)-\psi^{(0)}\left(\frac{s}{2}\right)\right)+
	\int_0^1 dx\ x^{s-1}\frac{\text{Li}_2(x)}{1+x}\Bigg].
\end{equation}
$\gamma\approx0.577215664901$ is the Euler-Mascheroni constant, $\psi^{(n)}$ are the polygamma functions
\begin{equation}
	\psi^{(n)}(z)=\left(\frac{d}{dz}\right)^{n+1}\ln \Gamma(z),
\end{equation}
and
\begin{equation}
	\Gamma(z)=\int_0^\infty t^{z-1}\exp(-t)dt.
\end{equation}
\subsection{Implementation of DGLAP energy scale integration}
\par The first method of solving the DGLAP equation is to integrate Eq. (\ref{eq:dglap}) in $t$ using the Euler method (i.e. $f(t+dt)\approx f(t)+dtf'(t)$) for ordinary differential equations (ODEs), and this is the approach chosen by Hirai \cite{hirai}. In our Python code, we allow for either log-scaled or linear-scaled sampling of the integration variable $z$, and estimate the integrals on the range $[x,1]$ using Simpson's rule. Alternatively, one may use the trapezoidal rule for integral estimation, or another drop-in replacement available in SciPy~\cite{scipy}. Practically, these choices do not make much difference in the numerical results, particularly if we choose a large number of integration points ($n_z\sim10^3$). We use NumPy to handle array manipulations~\cite{numpy}.
\par The Python code is a small set of modules which may be used in command line or imported as a package.
\subsection{Implementation of the DGLAP moment method}
The second method of solving the DGLAP equations is to perform the Mellin convolution in Mellin space and invert the result in Eq. (\ref{eq:moment}). While estimating the Mellin moment is easy, performing the inverse operation is numerically challenging. Fortunately, fast approximations for the closely-related inverse Laplace transform have been proposed~\cite{cohen}:
\begin{equation}
	\mathcal{L}[f](s)=\int_0^\infty f(t)\exp(-st)dt
\end{equation}
The Mellin transform is simply a Laplace transform with the substitution $x=\exp(-t)$, and therefore the inverse Mellin transform can be expressed in terms of the inverse Laplace transform as:
\begin{equation}
	\mathcal{M}^{-1}[f](x)=\mathcal{L}^{-1}[f](-\ln x).
\end{equation}
We used the mpmath Python package~\cite{mpmath} which implements Cohen's method for fast Laplace inversion~\cite{cohen}. We also include an implementation of Cohen's method in our Mathematica script\footnote{The implementation is described at \href{https://gnpalencia.org/blog/2022/invertlaplace/}{https://gnpalencia.org/blog/2022/invertlaplace/}}. Cohen's method replaces the Mellin inversion formula with an accelerated alternating series, whose accuracy depends on the degree of approximation ($d_{\text{approx}}$) desired.
\section{Results}
\subsection{Numerical correctness of the transversity splitting function moments}
In Mathematica 14.1, we verified numerically that the Mellin moments of LO and NLO splitting functions in Eqs. (\ref{eq:LO}) and (\ref{eq:NLO}) match the expressions given by Vogelsang in Eqs. (\ref{eq:LO_mom}) and (\ref{eq:NLO_mom}). For example, for $N_c=3$ and $N_f=5$, we found that the relative error between numerical moments and the analytic moments was at most 0.15\%. We also implemented both Hirai's and Vogelsang's methods in Mathematica to check for correctness, although the performance is lacking.
\subsection{Solving the DGLAP equation}
\begin{figure}
	\centering
	\includegraphics[width=\linewidth]{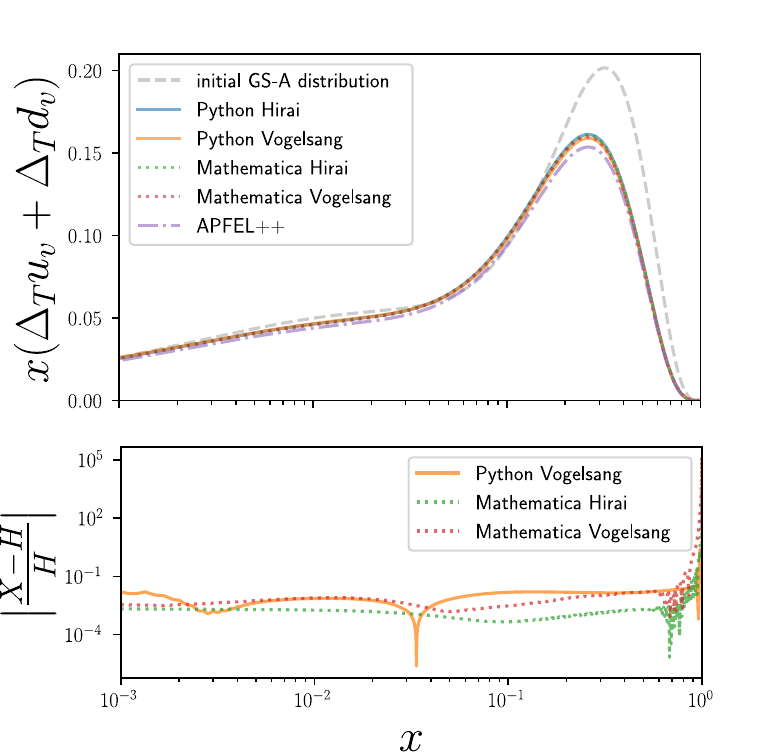}
	\caption{The GS-A distribution for $\Delta_T u_v+\Delta_T d_v$, evolved from 4 GeV$^2$ to 200 GeV$^2$ using both the Hirai method and the Vogelsang method at NLO. We also include the result of APFEL++, which was supplied by V. Bertone. The difference between our results and that of APFEL++ can be explained by different choices of $\alpha_S$, as discussed in the text. The bottom panel shows the absolute value of the difference between the other results and the benchmark, divided by the benchmark, where the benchmark is taken as ``Python Hirai''.}
	\label{fig:1}
\end{figure}
To verify the correctness of our implementations against Hirai's results, we used the NLO fitted Gehrmann-Stirling A-type longitudinally polarized distribution for the transversity PDFs of the up and down quarks at $Q^2=4\text{ GeV}^2$, $x\Delta_T q(x, Q^2)=x\Delta q(x, Q^2)$, with form given in \cite{gehrmann}:
\begin{equation}
	x\Delta_T q_v(x, Q^2)=\eta_q A_q x^{a_q}(1-x)^{b_q}(1+\gamma_q x+\rho_q \sqrt x),
\end{equation}
where $q$ is $u$ or $d$,
\begin{equation}
	A_q^{-1}=\bigg(1+\gamma_q \frac{a_q}{a_q+b_q+1}\bigg)\frac{\Gamma(a_q)\Gamma(b_q+1)}{\Gamma(a_q+b_q+1)}+\rho_q \frac{\Gamma\left(a_q+\frac12\right)\Gamma(b_q+1)}{\Gamma\left(a_q+b_q+\frac32\right)},
\end{equation}
$\eta_u=0.918,\eta_d=-0.339,a_u=0.512,a_d=0.780,b_u=3.96,b_d=4.96,\gamma_u=11.65,\gamma_d=7.81,\rho_u=-4.60,\rho_d=-3.48$. We evolved the minus type distribution $x(\Delta_T u_v+\Delta_T d_v)$ from 4 GeV$^2$ to 200 GeV$^2$, using the same settings as Hirai ($N_f=4$, $\Lambda_{QCD}=0.231$ GeV).
\par In Fig.~\ref{fig:1}, we present the results of our evolution package using both the Hirai~\cite{hirai} and Vogelsang~\cite{Vogelsang97} methods, as well as both Python and Mathematica programming languages. In the bottom panel of Fig.~\ref{fig:1}, we show that our results agree across different theoretical methods and different programming languages, with a relative difference on the order of 1\%. Additionally, we benchmark our results against that of APFEL++, where a numerical discrepancy is observed between our results and theirs, the cause of which we discuss in Sec.~\ref{sec:dis}. For the Hirai method, both Python and Mathematica results are shown with $N_t=100$.  The Mathematica version of the Hirai method was performed on $N_x=300$ points, due to the slowness of the implementation, while the Python one used $N_x=3000$. The blue solid line in our Fig.~\ref{fig:1} matches with Fig. 5 of Hirai~\cite{hirai}.  While the Vogelsang method does not discretize in $t$ space, the degree of approximation $d_{\text{approx}}$ for the inverse Mellin transform must be taken $d_{\text{approx}}\geq 5$ in order to obtain an accurate result, while no additional benefit in accuracy is seen if it is taken beyond $7$ (see Sec.~\ref{sub:per}).
\begin{figure}
	\centering
	\vspace{10pt}
	\includegraphics[width=\linewidth]{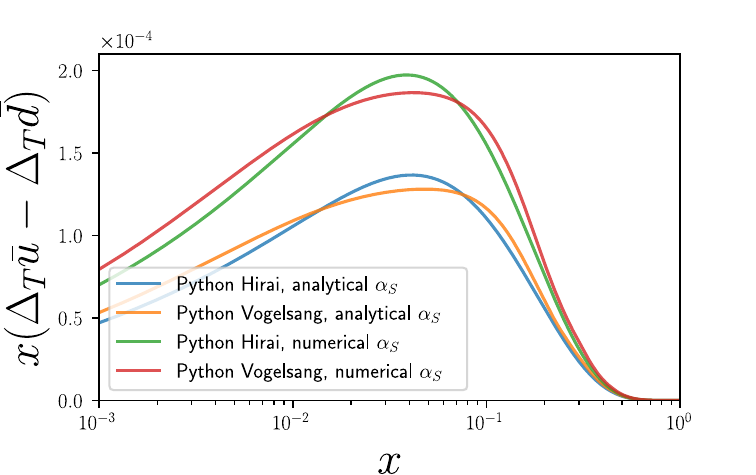}
	\caption{The GS-A distribution for $x(\Delta_T \bar u-\Delta_T \bar d)$, evolved from 4 GeV$^2$ to 200 GeV$^2$ using both the Hirai method and the Vogelsang method, as well as both choices of $\alpha_S$ at NLO. See Sec.~\ref{sec:dis} for a discussion on different choices of $\alpha_S$. Cohen's method is used to degree 5 in both the Vogelsang curves, without much improvement in agreement at higher degrees (not shown).}
	\label{fig:2}
\end{figure}
\begin{figure}
	\centering
	\includegraphics[width=\linewidth]{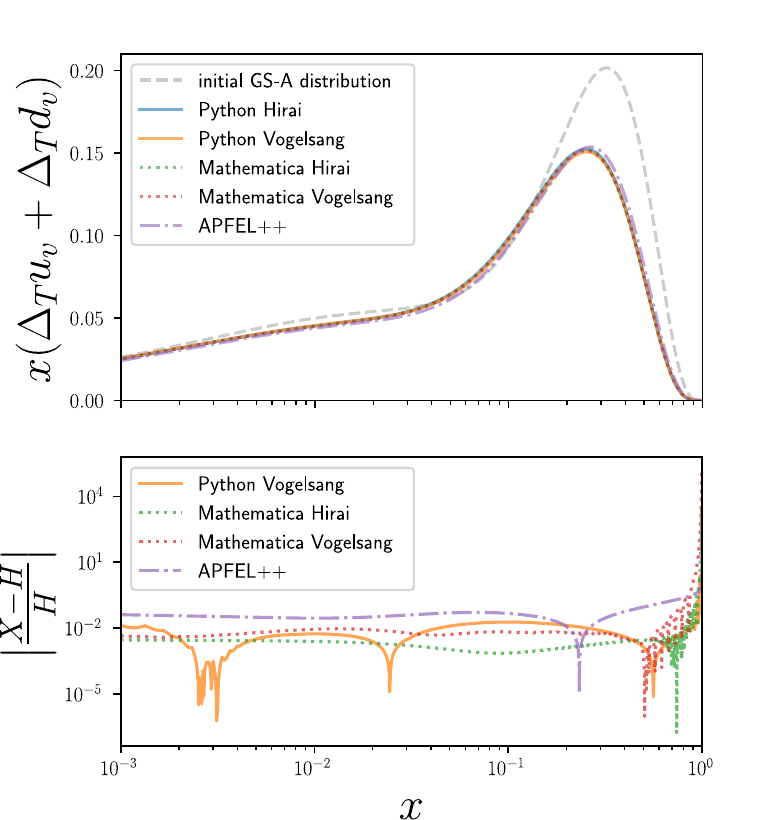}
	\caption{The GS-A distribution for $\Delta_T u_v+\Delta_T d_v$, evolved from 4 GeV$^2$ to 200 GeV$^2$ using both the Hirai method and the Vogelsang method at NLO, however with $\alpha_S$ being the numerical solution to the NLO $\alpha_S$ evolution equation, rather than the approximate analytical formula given in both Eq. (A.2) of Ref.~\cite{hirai} and Eq. (21) of Ref.~\cite{Vogelsang97}. The result of APFEL++ is the same as in Fig.~\ref{fig:1}. The bottom panel shows the absolute value of the difference between the other results, including the APFEL++ result, and the benchmark, divided by the benchmark, where the benchmark is taken as ``Python Hirai'' (with numerical $\alpha_S$).}
	\label{fig:1_exact_alpha}
\end{figure}
\par Similarly, we evolved the plus type distribution $x(\Delta_T u^+-\Delta_T d^+)=x(\Delta_T u_v-\Delta_T d_v)$ since $\Delta_T \bar{u}=\Delta_T \bar{d}$, and the minus-type distribution $x(\Delta_T u_v-\Delta_T d_v)$, and computed $x(\Delta_T \bar u-\Delta_T \bar d)=\frac x2\left((\Delta_T u^+-\Delta_T d^+)-(\Delta_T u_v-\Delta_T d_v)\right)$ in Fig.~\ref{fig:2}. The blue solid line in our Fig.~\ref{fig:2} matches Fig. 6 of Hirai~\cite{hirai}. For the Vogelsang method, the results in Fig.~\ref{fig:2} were again achieved with the degree of approximation equal to 5, and we note that further increasing the degree worsens the result. There is a relative discrepancy (the absolute error is a few parts in $10^5$) even between the Mathematica vs Python implementations of Hirai's method, likely due to numerical error.
\par In both Fig.~\ref{fig:1} and Fig.~\ref{fig:2}, we omit the LO evolution, however our code contains the capability to perform LO evolution using both Hirai's method and Vogelsang's method by adjusting the \verb|order| parameter.
\section{Discussion}
\subsection{Discrepancy between the evolution results\label{sec:dis}}
There exist discrepancies between our results and that of other works, for example, APFEL++, as well as between the two methods employed by us, which are due to a combination of two reasons. The first reason is that the analytical expression of $\alpha_S$ given in Eq. (\ref{eq:alpha}), which was used by both Refs.~\cite{Vogelsang97} and~\cite{hirai}, is only an approximation, and does not satisfy the NLO evolution equation of $\alpha_S$ exactly, especially in the smaller $Q^2$ region. We posit that this is the main source of the discrepancy between our results and that of APFEL++ observed in Fig.~\ref{fig:1}. In our code, we provide additional functionality to let the user choose whether to use the approximate formula Eq. (\ref{eq:alpha}), or to evolve $\alpha_S$ numerically starting from some reference scale, e.g. $\alpha_S(Q^2=M_Z^2)=0.118$. This numerical evolution is accomplished according to the LO or NLO evolution equation for the strong coupling constant $\alpha_S$, as given in Eq. (2.1) of Ref.~\cite{qcdnum}. In Fig.~\ref{fig:1_exact_alpha}, we plot the results if the user were to choose this numerical NLO evolution of $\alpha_S$ rather than Eq. (\ref{eq:alpha}), benchmarked against that of APFEL++. APFEL++'s result is shown twice without change in Figs.~\ref{fig:1} and ~\ref{fig:1_exact_alpha}; the difference between these two figures shows that the choice of $\alpha_S$ causes a significant change in the evolution, even when using high energy scales (from 4 GeV$^2$ to 200 GeV$^2$). This is despite the fact that the results in Figs.~\ref{fig:1} and ~\ref{fig:1_exact_alpha} are both valid and perturbatively equivalent to each other.
\par The second reason for the discrepancy between Hirai's method and Vogelsang's method is instability in the numerical Mellin inversion. For this reason, in both Figs.~\ref{fig:1} and \ref{fig:1_exact_alpha}, we can see that the relative difference between different methods is larger than the difference between programming languages. The instability in the numerical Mellin inversion is a well-known problem with both the inverse Laplace and Mellin problems due to the exponential decay of signal at $s\rightarrow \infty$ for the Laplace transform and $s\rightarrow 0$ for the Mellin transform; Cohen's method merely approximates the inversion formula Eq. (\ref{eq:mellininversion})~\cite{cohen}. The degree of approximation in Cohen's method determines the size of the discrepancy between the two methods, as seen in Fig.~\ref{fig:3_optimal_degree}. Again, the results in Figs.~\ref{fig:1}, \ref{fig:2}, and \ref{fig:1_exact_alpha} are all obtained with the degree of approximation equal to 5.
\subsection{Performance and accuracy\label{sub:per}}

\begin{figure}
	\centering
	\includegraphics[width=\linewidth]{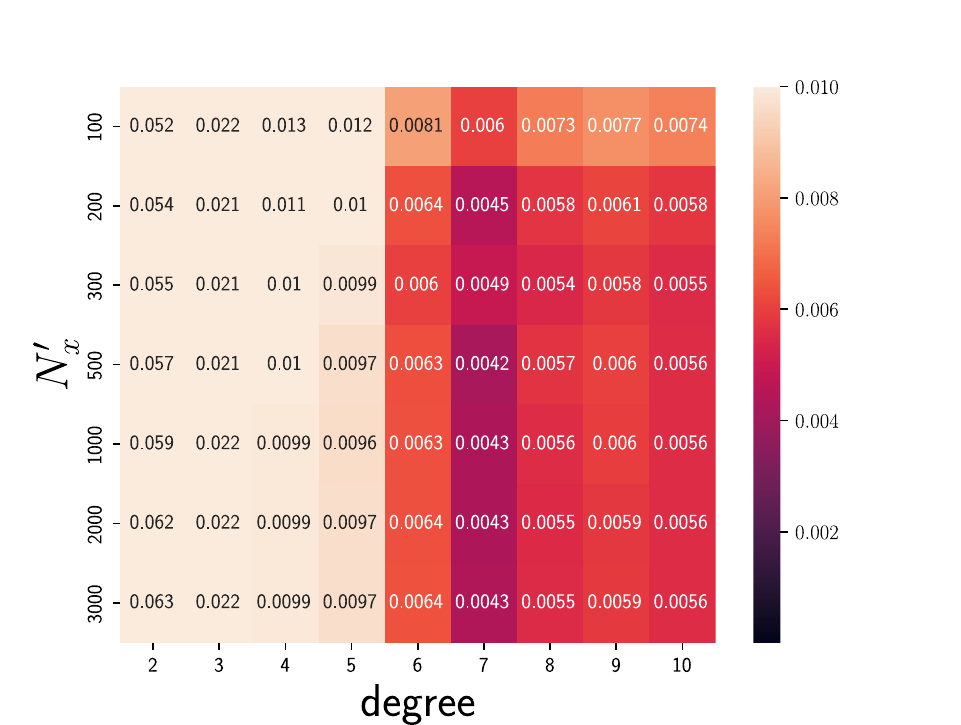}
	\caption{The relative error in computing Fig.~\ref{fig:1} as the degree of approximation and input PDF granularity $N_x'$ is varied in the Vogelsang method. The error is measured by the absolute value of the total area under the curve for the Vogelsang method minus the reference, divided by the area of the reference, where the reference is taken as ``Python Hirai" in Fig.~\ref{fig:1}.}
	\label{fig:3_optimal_degree}
\end{figure}

The Python codes perform evolution within a few minutes on an M3 MacBook Air. Since the number of flavors and colors contribute only to the various theory constants and not to the computational work, the time complexities of both methods are independent of these attributes. Note that we implement fixed flavor evolution.
\par In both methods, the input is a PDF sampled numerically at $N_x'$ points, and subsequently interpolated for further integration; $N_x'$ is independent of the user controlled $N_x$. In the Hirai method, the PDF is evolved through $N_t$ time steps, where at each time step, the incrementally evolved PDF is sampled numerically and interpolated at $N_x$ points rather than $N_x'$, giving rise to an estimated $O(N_x'+N_xN_t)$ complexity for Hirai. For Vogelsang, each of the $N_x$ points is computed independently from the other points, requiring the evaluation of the moments of the interpolated PDF of $N_x'$ points at a constant number of points (equal to $d_{\text{approx}}$) in the complex plane. $d_{\text{approx}}$ describes the number of terms in the alternating series are summed. Therefore the time complexity is $O(N_x' N_x d_{\text{approx}})$. Note that the accuracy of Hirai is dependent on $N_x$, whereas the accuracy of Vogelsang is independent of $N_x$. The accuracy of both methods is dependent on $N_x'$, i.e. the number of interpolation points in the provided numerical PDF.
\par
\begin{figure}
	\centering
	\includegraphics[width=\linewidth]{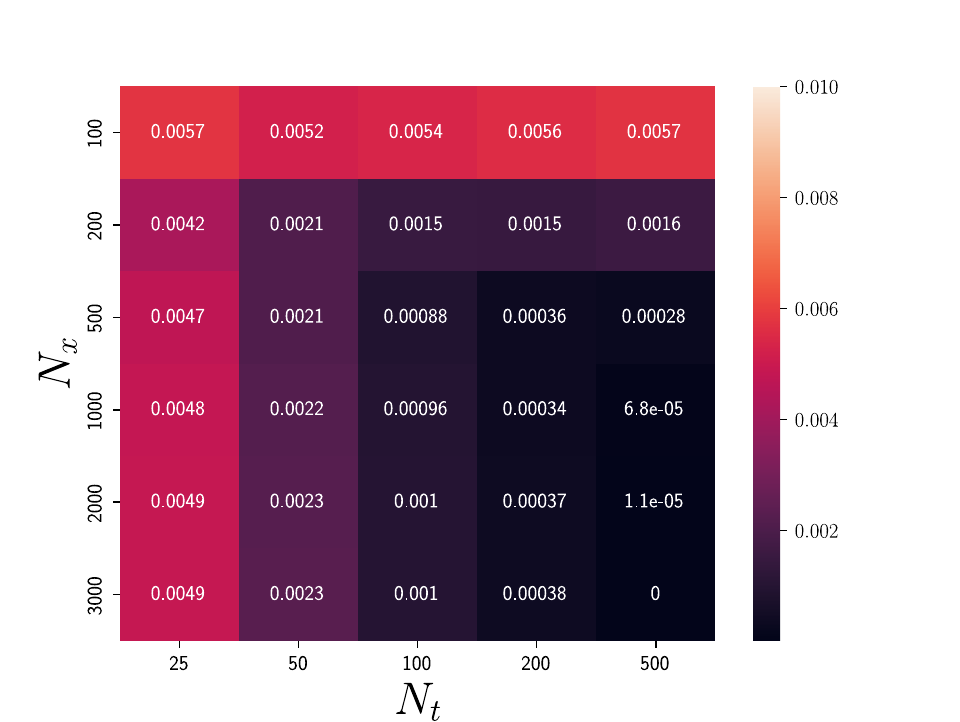}
	\caption{Relative error in computing Fig.~\ref{fig:1} as $N_x$ and $N_t$ are varied in the Hirai method. The error is measured by the absolute value of the total area under the curve for the Hirai method with sub-optimal $N_x$ and $N_t$, minus the benchmark, divided by the benchmark, where the benchmark is taken as the total area under the curve of the Hirai method with the $N_x=3000$ and $N_t=500$. This is done with Python. Note that the colorscale is the same as in Fig.~\ref{fig:3_optimal_degree}.}
	\label{fig:4_heatmap}
\end{figure}

 In Fig.~\ref{fig:3_optimal_degree}, we show a heatmap of the relative errors by varying $d_{\text{approx}}$ and $N_x'$ for the Vogelsang method, and in Fig.~\ref{fig:4_heatmap}, we show an analogous heatmap varying $N_x$ and $N_t$ for the Hirai method. In both Fig.~\ref{fig:3_optimal_degree} and Fig.~\ref{fig:4_heatmap}, the $N_x=3000$ and $N_t=500$ Hirai-evolved curve was used as the ground truth $f_{\text{ref}}$, and the relative error of $f(x)$ was defined as:
\begin{equation}
	\text{relative error}=\frac{\int_0^1 |f(x)-f_{\text{ref}}(x)|\ dx}{\int_0^1 |f_{\text{ref}}(x)|\ dx}.
\end{equation}
The relative error has a global minimum at degree $7$ and $N_x'=500$ for Vogelsang, whereas Hirai converges more uniformly for increasing $N_x$ and $N_t$. To achieve 1\% relative error while achieving optimal performance, we recommend $5\leq d_{\text{approx}}\leq 7$ and $N_x'\geq 200$ when using Vogelsang 's method. To achieve 0.1\% relative error, we recommend $N_x\geq 200$, $N_t\geq 100$ when using Hirai's method, in accordance with Figs.~\ref{fig:3_optimal_degree} and~\ref{fig:4_heatmap}. On an M3 MacBook Air, both methods at these settings took under a minute. Both methods allow for evolution from $Q_0$ to higher or lower $Q$. At high values of $N_x$ ($\sim 2000$), Hirai's method does outperform Vogelsang's. However, if one wishes only for the evolved value of the transversity PDF at a specific $x$, Vogelsang will outperform Hirai by orders of magnitude, so there may be specific use cases where Vogelsang is preferred. The Mathematica evolution code for Hirai's method is much slower ($\sim10\times$ to $100\times$) than the equivalent Python code, though we did not attempt targeted optimization of the Mathematica code since it was used simply to demonstrate correctness.
\subsection{Conclusion}
In this work, we provide both Mathematica and Python code which implements the evolution of transversity parton distribution functions up to NLO. We have shown that our implementation matches that of Hirai~\cite{hirai}. Furthermore, we make available an alternative Mellin moment method for performing the evolution in both Mathematica and Python, and we show that the formulae are free of errors using Mathematica. This manuscript is self-contained, including all the equations needed to implement these methods.
\section{Code and data availability statement}
\verb|tParton| is available on the Python Package Index at \url{https://pypi.org/project/tparton/} and on GitHub at \url{https://github.com/mikesha2/tParton/}, and may be installed on most Python-capable computers with \verb|pip| or the \verb|conda| package manager. A copy of tParton as well as the Jupyter and Mathematica notebooks for reproducing this paper can be found on Zenodo (doi: 10.5281/zenodo.17634737).
\section{Acknowledgments}
We have no funding sources which directly supported this work. We have no conflicts of interest to disclose. We thank Ian Clo\"et for helpful discussions regarding transversity PDF evolution. We also thank Valerio Bertone for supplying the results of APFEL++ evolution in Figs.~\ref{fig:1} and \ref{fig:1_exact_alpha}. CMS is grateful for support by the Penn State College of Medicine's Medical Scientist Training Program.

%% The Appendices part is started with the command \appendix;
%% appendix sections are then done as normal sections
%% \appendix

%% \section{}
%% \label{}

%% References
%%
%% Following citation commands can be used in the body text:
%% Usage of \cite is as follows:
%%   \cite{key}         ==>>  [#]
%%   \cite[chap. 2]{key} ==>> [#, chap. 2]
%%

%% References with BibTeX database:

\bibliographystyle{elsarticle-num}

\end{document}